\documentclass[aps,twocolumn,prl,showpacs,superscriptaddress]{revtex4-1}

\usepackage{graphicx}
\usepackage{bm}
\usepackage{hyperref}
\usepackage{color}
\usepackage{amsfonts}
\usepackage{tikz}
\usepackage{amsmath}

\newcommand{\St}{St}

\newcommand{\N}{n}
\renewcommand{\Re}{\mbox{\it Re}}
\newcommand{\K}{\lambda^\infty}

\newcommand{\ocaaddress}{Laboratoire J.-L.\ Lagrange, Universit\'e
  C\^ote d'Azur, OCA, CNRS, Bd.\ de l'Observatoire, 06300 Nice,
  France}
\newcommand*\circledi{{\tikz[baseline=(j.base)]{
            \node[shape=circle,draw,inner sep=0.5pt] (j)
            {\small$\phantom{j}\hspace{-4pt} i$};}}}
\newcommand*\circledj{{\tikz[baseline=(j.base)]{
            \node[shape=circle,draw,inner sep=0.5pt] (j)
            {\small$j$};}}}
\newcommand*\circledk{{\tikz[baseline=(j.base)]{
            \node[shape=circle,draw,inner sep=0.5pt] (j)
            {\small$\phantom{j}\hspace{-4pt} k$};}}}
\newcommand*\circledone{\tikz[baseline=(j.base)]{
            \node[shape=circle,draw,inner sep=0.5pt] (j)
            {\small$\phantom{j}\hspace{-4pt} 1$};}}

\begin{document}
%%%
\title{Abrupt growth of large aggregates by correlated coalescences in
  turbulent flow}

\author{J\'er\'emie Bec}
\affiliation{\ocaaddress}
\author{Samriddhi Sankar Ray}
\affiliation{International Centre for
  Theoretical Sciences, Tata Institute of Fundamental Research,
  Bangalore 560012, India}
\author{Ewe Wei Saw}
\affiliation{\ocaaddress}
\affiliation{Laboratoire SPHYNX, SPEC, CEA Saclay, CNRS, 91191 Gif-sur-Yvette, France}
\author{Holger Homann}
\affiliation{\ocaaddress}

\begin{abstract}
  Smoluchowski's coagulation kinetics is here shown to fail when the
  coalescing species are dilute and transported by a turbulent
  flow. The intermittent Lagrangian motion involves correlated violent
  events that lead to an unexpected rapid occurrence of the largest
  particles. This new phenomena is here quantified in terms of the
  anomalous scaling of turbulent three-point motion, leading to
  significant corrections in macroscopic processes that are critically
  sensitive to the early-stage emergence of large embryonic
  aggregates, as in planet formation or rain precipitation.
\end{abstract}


\maketitle
\noindent The formation of planets in circum-stellar
disks~\cite{weidenschilling1993formation,lissauer1993planet} as well
as the initiation of rain in warm
clouds~\cite{pinsky1997turbulence,shaw2003particle} involve the
coalescence of small dilute bodies suspended in a highly turbulent
flow.  It is crucial, in both cases, to determine the speed at which
the largest objects are formed.  Massive planetary embryos or big
raindrops decouple from the underlying flow and accrete smaller
particles more
efficiently~\cite{wetherill1989accumulation,kostinski2005fluctuations,johansen2015growth}.
They are, very likely, the precursors for a run-away growth and
possibly trigger the full coagulation process. Turbulent fluctuations
might be essential in the formation of such large
objects~\cite{johansen2007rapid,bodenschatz2010can} but their precise
role is still far from being fully understood.  Significant progress
has been made in understanding the enhancement of kinetic collision
kernels due to turbulence. It is important to recall two key
mechanisms present in the particle dynamics: \emph{preferential
  concentration}~\cite{sundaram1997collision}, giving rise to high
densities, and the \emph{sling
  effect}~\cite{falkovich2002acceleration} or \emph{caustic
  formation}~\cite{wilkinson2006caustic}, responsible for large
velocity differences; both mechanisms enhance the rate at which
particles approach each other. Precise quantitative models accounting
for these two effects require appreciating the influence of
turbulence~\cite{devenish2012droplet,pan2014turbulence}. However,
their origin is not directly related to turbulent fluctuations but
rather comes from the inertia of the suspended particles and the
resulting detachment of their trajectories from the flow. Their impact
on collision rates can then be studied in simple random
flow~\cite{bec2005clustering,gustavsson2013distribution,vosskuhle2014prevalence}.

In this Letter we show that, by its own, turbulent transport speeds up
the growth of large objects.  In the Lagrangian evolution of fluid
elements, scaling and geometry are tied up by non-trivial memory
effects.  These interdependences lead to \emph{intermittent
  multiscaling} properties of advected passive scalar
fields~\cite{shraiman2000scalar,falkovich2001particles}.  In the
context of growth by coagulation, as shown in this work, they are
responsible for a power-law tail in the distribution of times between
successive collisions, yielding intricate correlations in the sequence
of coalescences experienced by individual particles. Because of this
effect, we find that the number of large objects grows as a power law
at short times, with an exponent much smaller than the one obtained
from kinetic population-balance approaches. The value of this exponent
is expressed in terms of the anomalous scaling exponent $\zeta_3$
associated to the third-order correlations of an advected passive
scalar.

To simplify the presentation, we focus on an initially mono-disperse
suspension consisting of $\N_1$ monomers $\circledone$ with mass
$m_1$.  The extension to poly-disperse situations is straightforward.
These particles evolve in a turbulent flow and might coalesce,
summing-up their masses, when they collide.  This dynamics leads,
after sometime, to the formation of a broad spectrum of particle
sizes. We denote by $\circledi$ those constituted of $i$ monomers and
thus with a mass $i\times m_1$. Our goal is to determine how fast the
number $\N_i(t)$ of particles $\circledi$ grows with time for $i >
1$. Simple population-balance considerations lead to
\begin{equation}
  %\frac{\mathrm{d}\,\N_i}{\mathrm{d}t} 
  \dot{\N}_i(t)= \frac{1}{2}
  \sum_{j=1}^{i-1}\mathcal{Q}_{i-j,j}(t) - \sum_{j=1}^\infty
  \mathcal{Q}_{i,j}(t),
  \label{eq:evol_numbers}
\end{equation}
where the dot denotes time derivative.
$\mathcal{Q}_{i,j}(t)\,\mathrm{d}t$ is the number of coalescences
$\circledi+\circledj$ occurring between times $t$ and
$t+\mathrm{d}t$. The first term in the right-hand side, the source,
accounts for the rate at which particles $\circledi$ are created. The
second, the sink, handles the coalescences of such particles with all
others. When $\N_i(0)=0$ (for $i > 1$), the global coalescence rate
$\mathcal{Q}_{i,j}$ can be written in terms of the individual particle
rate by summing over all the creations of $\circledi$'s
\begin{equation}
  \mathcal{Q}_{i,j}(t) = \int_0^t \lambda_{i,j}(t-s|s)\, \N_j(t)
  \,\dot{\N}_i(s) \, \mathrm{d}s.
  \label{eq:Qij}
\end{equation}
$\lambda_{i,j}(\tau|s)$ is the rate at which a particle $\circledi$,
created at time $s$, coalesce with a $\circledj$ at time $s+\tau$. For
statistically steady particle dynamics, this quantity is independent
of the creation time $s$ and
$\lambda_{i,j}(\tau|s) = \lambda_{i,j}(\tau)$. Also, this rate relates
to the probability distribution $p_{i,j}(\tau)$ of the time to next
collision, which is given by
\begin{equation}
  p_{i,j}(\tau) = \lambda_{i,j}(\tau)\, \mathrm{e}^{-\int_0^{\tau}
    \lambda_{i,j}(\tau') \, \mathrm{d}\tau'}.
  \label{eq:distrib_intercol}
\end{equation}
This is the distribution of waiting time associated to the
non-homogeneous Poisson process with rate $\lambda_{i,j}(\tau)$.

At sufficiently long times $\tau$, the coalescence rate
$\lambda_{i,j}(\tau)$ is expected to approach a finite limit
$\K_{i,j}$.  Successive collisions of a single particle then appear to
be uncorrelated. They define a memoryless process and $p_{i,j}(\tau)$
tends to the exponential distribution with rate parameter $\K_{i,j}$. The
population-balance system (\ref{eq:evol_numbers})-(\ref{eq:Qij}) then
reduces to
\begin{equation}
  \dot{\N}_i = \frac{1}{2} \sum_{j=1}^{i-1}
  \K_{i-j,j}\,\N_{i-j} \, \N_j -  \sum_{j= 1}^\infty
  \K_{i,j}\,\N_i \, \N_j.
  \label{eq:smolu}
\end{equation}
This is the celebrated Smoluchowski coagulation
equation~\cite{smoluchowski1917versuch}. The stationary rates
$\K_{i,j}$ are usually referred to as the collection or coalescence
kernels.  The work cited above on particle inertia was actually
devoted to estimating their dependence upon particle sizes and the
turbulent fluctuations of the carrier flow. The kinetic
model~(\ref{eq:smolu}) leads to predictions concerning the short-time
increase of the $\N_i$'s.  At the early stages of particle growth, the
number $\N_1$ of monomers remains almost constant and creations are
dominant in the population balance. We thus get
$\dot{\N}_2 \simeq \K_{1,1}\, \N_1^2/2$, so that
$\N_2(t) \simeq \N_1^2\, \K_{1,1}\,t /2$.  For the next size, we have
$\dot{\N}_3 \simeq \K_{1,2}\, \N_1\,\N_2$ and thus
$\N_3(t) \simeq \N_1^3\, \K_{1,1}\,\K_{1,2}\,t^2 /4$. We obtain
recursively
\begin{equation}
  \N_i(t) \simeq \N_1^i \left({t}/{t_i}\right)^{i-1},
  \label{eq:evol_mf}
\end{equation}
where the times $t_i$ are averages of the times $1/\K_{j,k}$
associated to the different combinations of coalescences
$\circledj + \circledk$ that are necessary to form a particle
$\circledi$.  The consistency of the assumptions can be checked a
posteriori: The creation terms in (\ref{eq:smolu}) are always
$\propto t^{i-2}$ and thus prevail at short times over the dominant
destruction term $\propto t^{i-1}$.

The main assumption leading to Smoluchowski kinetics (\ref{eq:smolu})
is a convergence of the coalescence rate to its limiting value
$\K_{i,j}$ much faster than the evolution of $n_i$.  This is ensured
for instance when the particles are very dense.  For explaining the
formation of large particles in a dilute suspension, these timescales
are in general not sufficiently separated. The sudden appearance of
sizable aggregates requires a brisk sequence of coalescences that are
very likely to be correlated to each other. When, in addition, the
coalescing species are transported by a turbulent flow, such
correlations speed up the growth of large particles.

A statistically steady turbulent flow involves interactions between
eddies of various sizes, ranging from the integral scale $L$, where
kinetic energy is injected at a rate $\varepsilon$, down to the
dissipative scale $\eta = \nu^{3/4}/\varepsilon^{1/4}$ below which
viscous damping dominates ($\nu$ denotes the kinematic viscosity of
the fluid).  The degree of turbulence grows with the extension of this
spatial span and is measured by the Reynolds number
$\Re = (L/\eta)^{3/4}$.  The intermediate scales between $\eta$ and
$L$ define the inertial range through which energy cascades with a
rate $\varepsilon$.  Dimensional arguments suggest that the velocity
increments between two points separated by a distance $r$ in the
inertial range behave as $u_r \sim (\varepsilon r)^{1/3}$.  Such a
phenomenology, referred to as Kolmogorov 1941, is often enough for
capturing the most significant effects of turbulent fluctuations; in
reality the scaling properties display slight deviations, due to
intermittency, from this dimensional
prediction~\cite{frisch1995turbulence,falkovich2006lessons}.

\begin{figure}[t!]

  \includegraphics[height=.6\columnwidth]{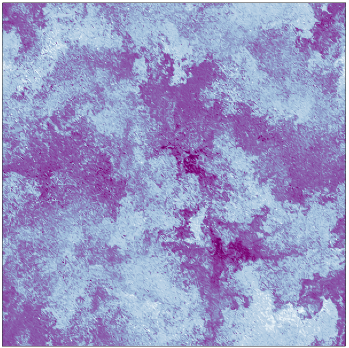}
  \hfill\includegraphics[height=.55\columnwidth]{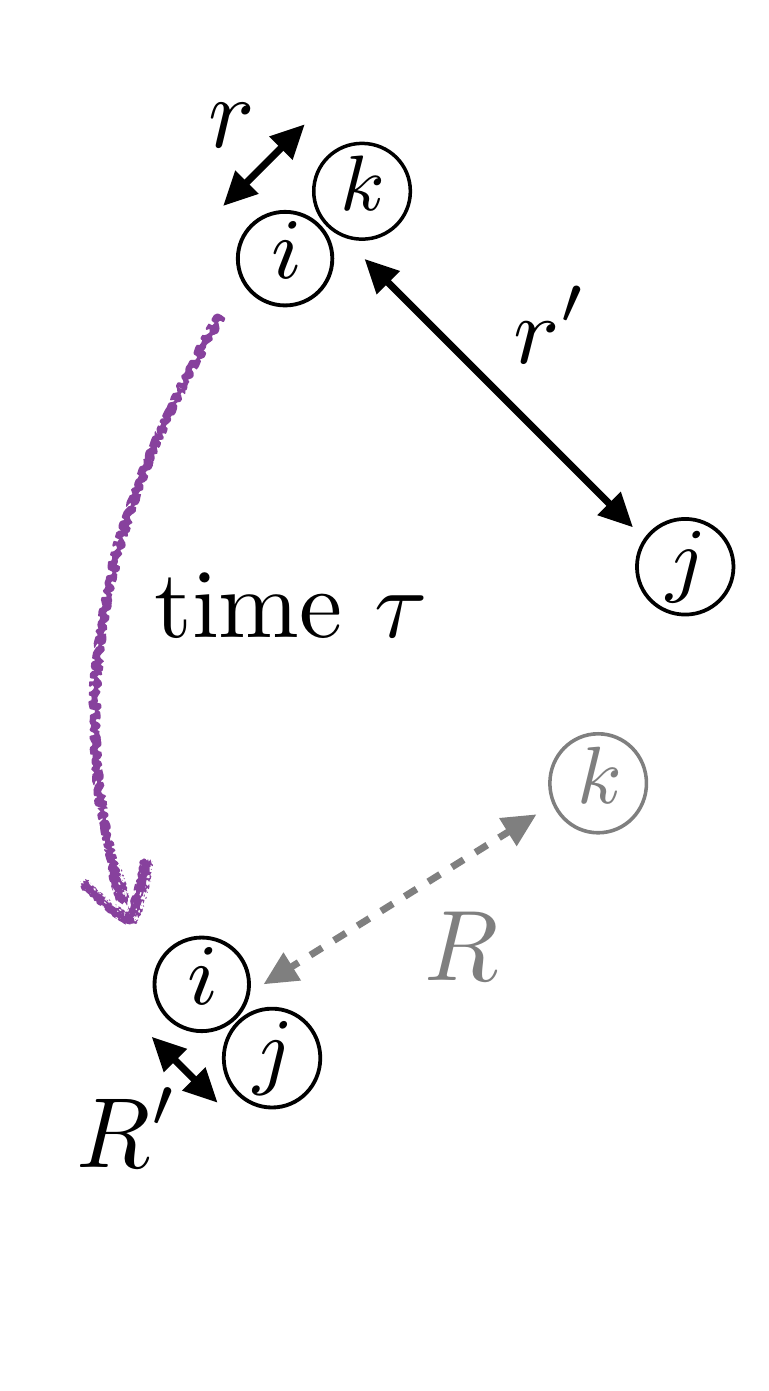}
  \vspace{-10pt}
  \caption{\emph{Left}: Distance traveled by fluid elements in a 3D
    turbulent flow during one large-eddy turnover time. Long (white) and
    short (purple) distances, represented here as a function of the
    final position in a 2D slice, define an intricate landscape with
    fronts where particles coming from far apart meet together.
    % There is a latex bug when the circles are put around the letters
    \emph{Right}: Sketch of the event leading to correlated successive
    collisions. At the initial time (top), two particles $i$ and $k$
    are located at a collision distance $r\lesssim\eta$, while a third
    one $j$ is at distance $r'\gg\eta$ far from them. A time $\tau$
    later (bottom) $j$ has approached $i$ at a distance
    $R'\lesssim\eta$ while $k$, having collided or not, has escaped to
    $R\gg\eta$.}
  \label{fig:snapshot}
\end{figure}

The breakdown of scale invariance is much more striking for mixing
statistics, owing to the fact that turbulence mingles together fluid
elements in a robust manner.  This pops up with the presence of
quasi-discontinuities in the Lagrangian map where materials
originating from distinct regions of the flow are violently brought
together (see Fig.~\ref{fig:snapshot} \emph{Left}).  The emergence of
such \emph{fronts} is due to the inertial-range roughness of the
velocity field and the associated non-uniqueness of fluid element
trajectories.  Two initially separate tracers $\bm x_1(t)$ and
$\bm x_2(t)$ that closely approach each other become indistinguishable
and separate afterwards following Richardson's superdiffusion
$|\bm x_1(t)-\bm x_2(t)|^2 \sim \varepsilon t^3$.  Still, when
interested in more than two fluid elements, this explosive behavior is
constrained by the underlying presence of \emph{statistical
  conservation laws} induced by the spatial correlations of the
velocity
field~\cite{shraiman2000scalar,falkovich2001particles,arad2001statistical}.
There exists specific functions of the shape and size of a cloud of
$n$ tracers that on average do not vary with time.  Such zero modes
are known to yield anomalous scaling in the statistics of an advected
passive scalar $\theta$.  Its structure functions behave in the
inertial range as
$\langle (\theta(\bm x+\bm r)-\theta(\bm x))^n\rangle \sim |\bm
r|^{n/3-\delta_n}$
where the discrepancies $\delta_n$ of the exponents from their
dimensional prediction relate to the anomalous scaling of the
transition probability of the distances between $n$ tracers. As we
will now see, the behavior of the three-point motion is in fact of
relevance to coalescences.

In dilute suspensions, a coalescence results from two successive
processes. First, the turbulent flow needs to bring two initially
separate particles at a sufficiently close distance $\lesssim\eta$.
Second, these close particles need to actually merge, and this
involves various microphysical mechanisms (particle inertia,
hydrodynamical interactions, surface effects). This leads us to write
the coalescence rate as a product of two contributions:
\begin{equation}
  \lambda_{i,j}(\tau) \approx \lambda_{i,j}^{\rm turb}(\tau) \times
  \lambda_{i,j}^{\rm micro}.
\end{equation}
The contribution from turbulent transport can be written
\begin{equation}
  \lambda_{i,j}^{\rm turb}(\tau) \approx \int  u_\eta \,
  p_3(R,\eta,\tau|\eta,r',0)\,
  ({r'^2}/{L^3})\,\mathrm{d}r'\mathrm{d}R
  \label{eq:turbrate_3p}
\end{equation}
involving the transition probability $p_3$ of the three-point
motion. More specifically, two successive collisions occur if three
particles (see Fig.~\ref{fig:snapshot} \emph{Right}), initially
separated by distances $r=\eta$ and an arbitrary $r'$, come in a time
$\tau$ to distances $R$, arbitrary, and $R'=\eta$.  The relation
(\ref{eq:turbrate_3p}) is obtained by integrating over all possible
initial distances $r'$ of the particle $\circledj$ from the particle
$\circledi$, with a weight $\propto r'^2$ given by a uniform
three-dimensional spatial distribution.  The three tracers
$\circledi$, $\circledj$ and $\circledk$ undergo in a time $\tau$ an
evolution from a degenerate triangle with $r\ll r'$ to another
degenerate triangle with, this time, $R\gg R'$. In turbulence, the
probability transition between such configurations can be written as
\begin{equation*}
  p_3(R,R'\!,\tau|r,r'\!,0) \approx \left(\frac{\eta}{r'}\right)^{\!2}\!
  \left(\frac{L}{r'}\right)^{\!\!\delta_3}\!\!
  \frac{1}{\varepsilon\tau^{3}}\Psi\!\left(\frac{r'^2}{\varepsilon\tau^{3}},
    \frac{R^2}{\varepsilon\tau^{3}}\right)\!,
\end{equation*}
where $\delta_3$ is the anomalous part of the scaling exponent
associated to the third-order statistics of an advected passive
scalar; its value is universal (independent of the injection
mechanism) and $\approx 0.18$, as reported from several experimental
and numerical studies~\cite{warhaft2000passive}.  In the expression
above the first factor comes from integration over angles and can be
seen as a small solid-angle contribution. The second factor originates
from intermittency and gives a dependence upon the integral scale
$L$. Physically, it means that when $\delta_3>0$, the closer is the
third particle, the more likely it is to approach one of the other
two.  The last terms involve a dimensionless function $\Psi$ that
imposes Richardson's scaling for backward and forward pair
evolution. This specific form of the three-point transition
probability leads to
\begin{align}
  \lambda_{i,j}^{\rm turb}(\tau)
  &\propto\frac{\nu^{7/4}}{\varepsilon^{1/4}L^{3-\delta_3}}\! \int
    \Psi\!\left(\frac{r'^2}{\varepsilon\tau^{3}},
    \frac{R^2}{\varepsilon\tau^{3}}\right)
    \frac{\mathrm{d}r'\mathrm{d}R}{r'^{\delta_3}\varepsilon\tau^3} \nonumber\\
  &\propto
    ({1}/{\tau_L}) \left({\tau}/{\tau_L}\right)^{-\frac{3}{2}\delta_3},
\end{align}
where $\tau_L= \varepsilon^{-1/3}L^{2/3}$ is the large-eddy
turnover time. Turbulent transport thus leads to a power-law
dependence in time of the coalescence rate. Plugging this behavior in
the global coalescence rate (\ref{eq:Qij}) and by using the
population-balance equations (\ref{eq:evol_numbers}), one obtains a
short-time behavior of the number of particles $\circledi$ that reads
\begin{equation}
  \N_i(t) \simeq \N_1^i \left({t}/{\tilde{t}_i}\right)^{(1-\frac{3}{2}\delta_3)(i-2)+1}.
  \label{eq:evol_turb}
\end{equation}
Here the characteristic times $\tilde{t}_i$ are $\propto i\,\tau_L$,
with a proportionality constant that involves the various
microphysical rates $\lambda_{j,k}^\text{micro}$ of the coalescences
leading to $\circledi$.  For $\delta_3>0$ the algebraic exponent
appearing in (\ref{eq:evol_turb}) is smaller than that obtained in
(\ref{eq:evol_mf}) from Smoluchowski's kinetics.  The intermittency of
turbulence mixing is thus enhancing the short-time growth by
coalescence. In addition, the larger is the aggregate size considered,
the stronger is this enhancement. Indeed, when $i$ is large and
$t\ll i\,\tau_L$, the formation of a particle $\circledi$ requires a
large number of correlated coalescences separated by inertial-range
times and the population dynamics is dominated by (\ref{eq:evol_turb}).

In order to corroborate our theoretical predictions on the enhancement of
coalescences by turbulent mixing, we have performed direct numerical
simulations for the evolution of a dilute population suspended in a
turbulent flow. We start from one billion inertial point-particles
whose dynamics is given by a viscous Stokes drag:
\begin{equation}
  \ddot{\bm x}_n = -\frac{1}{\tau_n} \left[\dot{\bm x}_n - \bm u(\bm
    x_n,t)\right]\!,
  \label{eq:stokes}
\end{equation}
where $\bm u$ designates the fluid velocity field. It is obtained
numerically by a pseudo-spectral integration of the incompressible
Navier--Stokes equation using $2048^3$ gridpoints. A large-scale
forcing is applied in order to maintain the flow in a developed
turbulent state with $\Re\approx 50\,000$.
 
The particles follow the flow with a time lag given by their
individual response times $\tau_n$.  Each particle has a virtual
radius $a_n(t)$ and $\tau_n\propto a_n^2$.  We start from a
mono-disperse suspension with monomers having initially all the same
radius $a_i(0) = a_0 \approx \eta/10$. When two particles approach at
a distance equal to the sum of their radii (detected using a billiard
algorithm), they merge, conserving mass and momentum. Particles
inertia is measured by their Stokes numbers
$\St_n=\tau_n \,\varepsilon^{1/2} / \nu^{1/2}$, which is initially
small $\St_n(0)\approx 0.1$. Inertia effects can thus be clearly
neglected when interested in inertial-range length or time scales. The
suspension is dilute: their volume fraction is approximately
$5\times10^{-5}$, which represents in our flow one particle for each
cube of volume $10\,\eta^3$ and is consistent with, for example,
typical settings in a warm cloud of our atmosphere.

\begin{figure}[ht]
  \includegraphics[width=0.9\columnwidth]{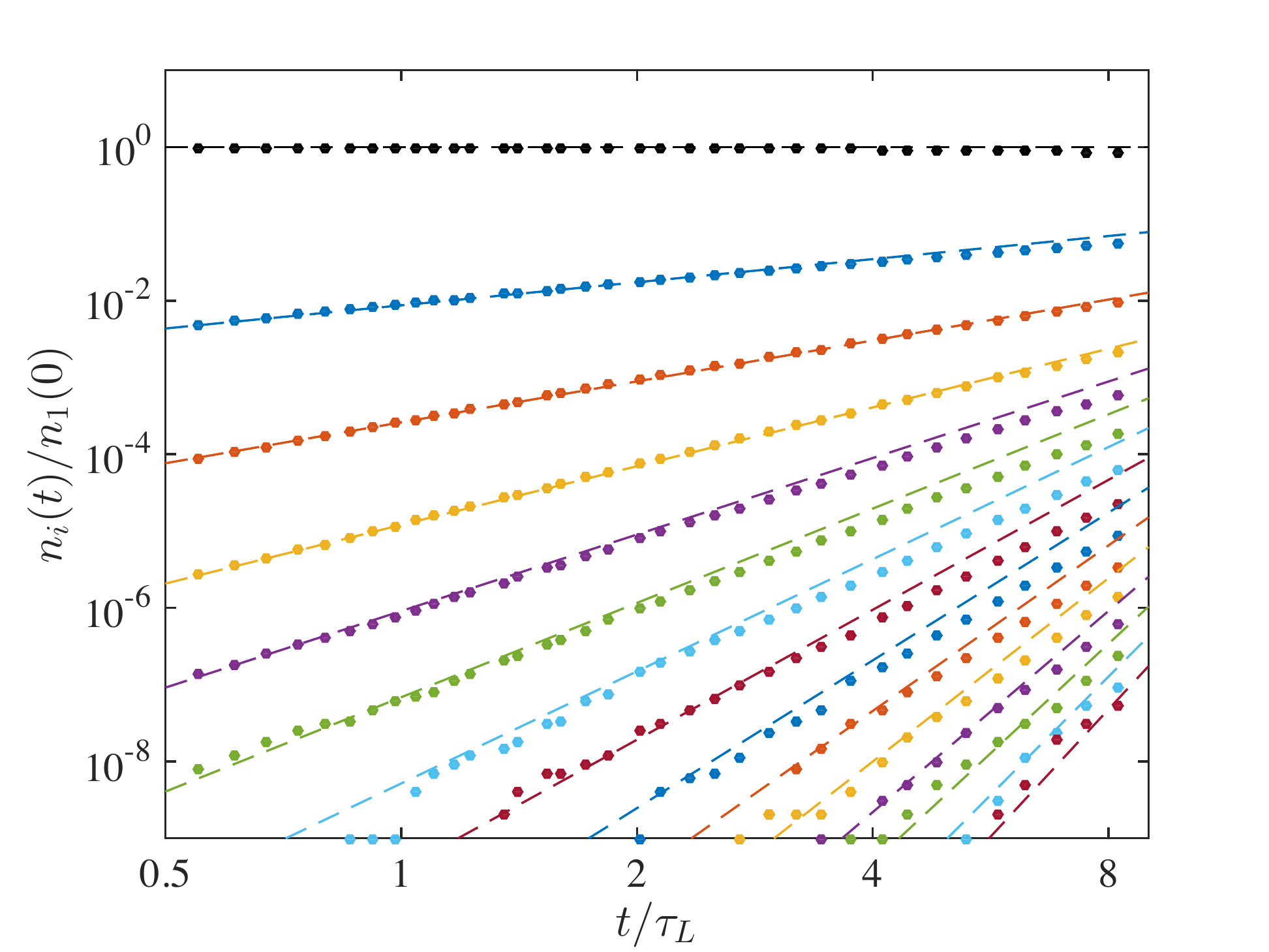}
  \vspace{-10pt}
  \caption{Time evolution of the number $\N_i(t)$ of particles
    (normalized by the initial number of monomers $\N_1(0)$); the mass
    $i$ increases from 1 to 15 from top to bottom. The dots are the
    results of direct numerical simulations and the dashed lines show
    for $i\ge 2$ behaviors $\propto t^{0.73(i-2)+1}$ deduced from
    (\ref{eq:evol_turb}) for $\delta_3 = 0.18$. }
  \label{fig:evol_npart}
\end{figure}
Figure~\ref{fig:evol_npart} shows on log-log scales the time evolution
of the number $\N_i(t)$ of particles made from the merger of $i$
monomers (that is with radius $a(t) = i^{1/3}a_0$) for
$i=1,2,\dots15$. Data (dots) is approximated very well by
$n_i(t) \propto t^{0.73(i-2)+1}$ (dashed lines), corresponding to the
predicted power laws (\ref{eq:evol_turb}) with $\delta_3 = 0.18$.
Such a behavior persists for times larger than the large-eddy turnover
time. The result of our simulation confirms the accuracy and the
relevance of the predictions made earlier in this Letter. Large
aggregates are appearing faster than predicted from kinetic models.

\begin{figure}[h]
  \includegraphics[width=0.9\columnwidth]{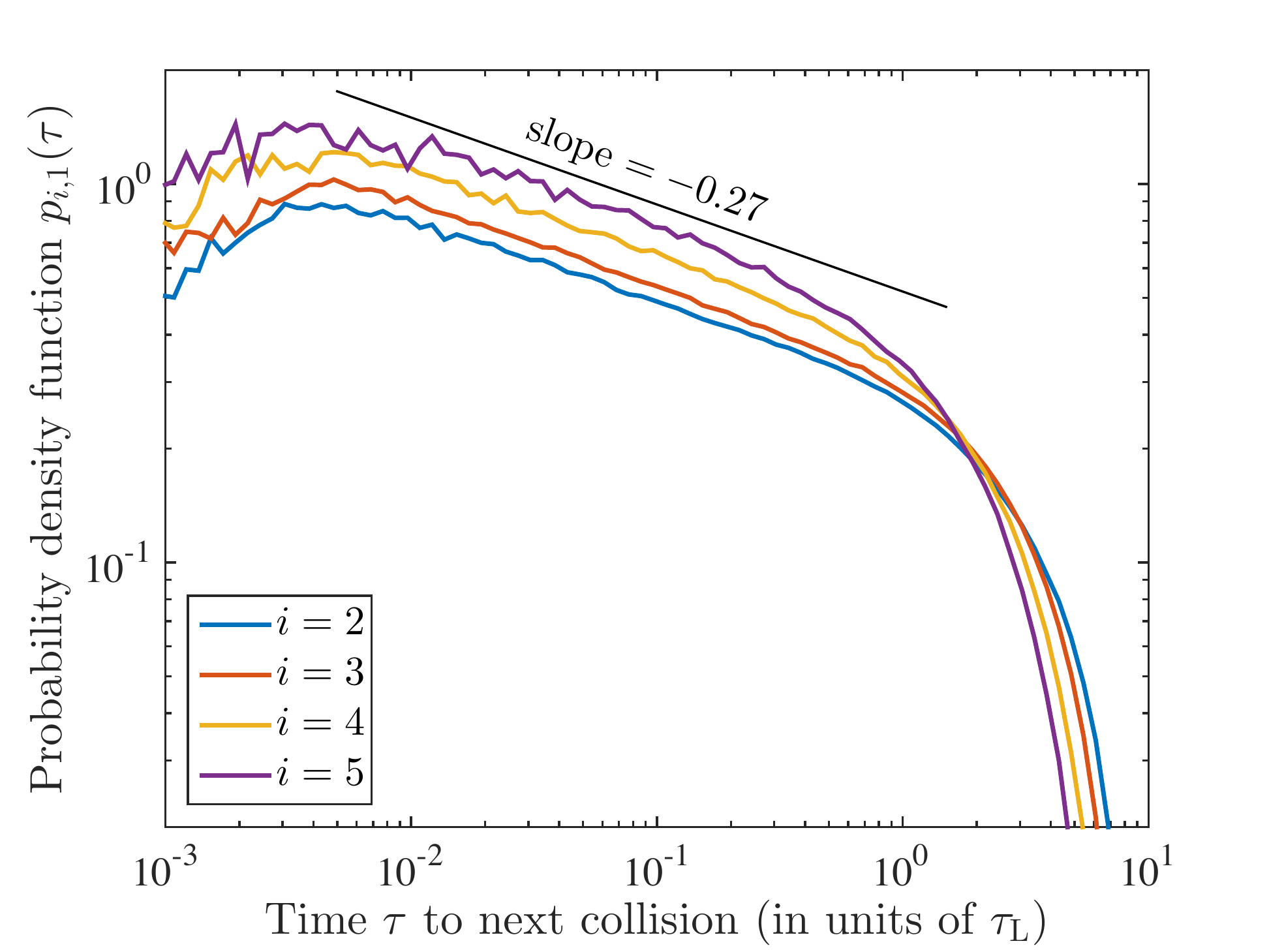}
  \vspace{-10pt}
  \caption{Probability density function of the time between successive
    coalescences of a particle $i$ with a monomer of mass $1$. The
    power-law tail has, as we predict theoretically, an exponent
    $\approx -0.27 \approx -({3}/{2})\delta_3$ }
  \label{fig:distrib_next_coll_time}
\end{figure}

To confirm that this enhanced growth is indeed resulting from
correlated successive collisions, we have measured the probability
density $p_{i,1}(\tau)$ of the time lag $\tau$ between the creation of
a particle $\circledi$ and its next collision with a $\circledone$.
Results are shown in Fig.~\ref{fig:distrib_next_coll_time} for
$i=2,3,4,5$. The distributions clearly display for $\tau$ in the
inertial range a power-law decay, followed by a (stretched)
exponential cutoff at $\tau\gtrsim \tau_L$. The measured value of the
algebraic exponent is consistent with the predicted value
$-(3/2)\delta_3\approx -0.27$.  This confirms that the inter-collision
time distribution follows (\ref{eq:distrib_intercol}) with a
time-dependent coalescence rate
$\lambda_{i.j}(\tau) \propto \tau^{-\frac{3}{2}\delta_3}$.

In conclusion, let us stress again that intermittency of turbulent
mixing is responsible for an enhanced growth of dilute coalescing
aggregates. To our knowledge, this is one of the first instances where
turbulent anomalous scaling laws play a critical role in describing to
leading order a process with practical implications. Here, only
third-order statistics are relevant since successive binary collisions
involves the evolution of triplets of tracers that form degenerate
triangles. Higher-order statistics enter other configurations (when,
for instance, two particles are simultaneously formed and then merge)
but they give subleading contributions. Finally, it is worth
mentioning that the effect unveiled here might be accounted for by
modifying kinetic models.  When coarse-graining the population
dynamics on sufficiently large timescales, correlated successive
collisions will then appear as simultaneous multiple collisions.

The research leading to these results has received funding from the
European Research Council under the European Community's Seventh
Framework Program (FP7/2007-2013, Grant Agreement no. 240579), from
the Agence Nationale de la Recherche (Programme Blanc
ANR-12-BS09-011-04). Access to the IBM BlueGene/P computer JUGENE at
the FZ J\"ulich was made available through the PRACE project
PRA031. JB and SSR were supported by the Indo-French Center for
Applied Mathematics (IFCAM). SSR acknowledges the AIRBUS Group for the
Corporate Foundation Chair in Mathematics of Complex Systems
established in ICTS-TIFR.

\bibliographystyle{apsrev4-1}
\bibliography{biblio}

\end{document}